\documentclass{optica-article}

\journal{opticajournal} % for journals or Optica Open

\articletype{Research Article}

\usepackage{lineno}
\usepackage{siunitx}
\usepackage[version=4]{mhchem}
\usepackage{booktabs} 
\usepackage{tabularx}
\usepackage{wasysym}

\begin{document}

\title{Characterization of Autofluorescence in Optical Fibers for NV-based Sensing Applications}

\author{Stefan Johansson,\authormark{1} Alexander Bukschat,\authormark{1} Dennis Lönard,\authormark{1} Alena Erlenbach, \authormark{1} Jonas Gutsche,\authormark{1} and Artur Widera\authormark{1*}}

\address{\authormark{1}Department of Physics and State Research Center OPTIMAS, RPTU University Kaiserslautern-Landau, Erwin-Schroedinger-Str. 46, 67663 Kaiserslautern, Germany}

\email{\authormark{*}widera@rptu.de}

\begin{abstract*}
Optical fibers are crucial for guiding light in various sensing applications.
Especially for quantum sensors such as the nitrogen-vacancy (NV) center in diamond, they enable light control and device miniaturization.
However, fluorescence and scattering within the fiber, often referred to as fiber background, autofluorescence, or autoluminescence, can overlap spectrally with the NV centers' fluorescence, degrading the signal-to-noise ratio and thus limiting sensor sensitivity.
Here, we investigate the optical spectra of standard optical fibers, considering material dependencies, physical influences, and their fluorescence scaling with excitation power and wavelength.
Our results identify spectral components and fiber types with minimal unwanted background signals, guiding the selection of optimal fibers for NV-based quantum sensing.
\end{abstract*}

%%%%%%%%%%%%%%%%%%%%%%%%%%  body  %%%%%%%%%%%%%%%%%%%%%%%%%%

\section{Introduction}
Optical fibers have become an indispensable element in today's world \cite{Winzer.2018}. 
They are particularly used in lighting, data transmission, and sensor technology \cite{Tembhare.31120203132020, Agrawal.2012, Bozinovic.2008, Elsherif.2022, Li.2022, Sabri.2015, Xin.2025, Flusberg.2005}.
Depending on the application, different optical fiber types are selected, e.g., for optimal transmission, mechanical robustness, or spectral characteristics.
In sensing, optical fibers are often utilized to guide light for time-resolved, intensity, or spectrally dependent measurements \cite{Ehrlich.2020, Bianco.2021, Quan.2023}.
Recently, fiber-coupled sensors based on fluorescence detection have gained interest due to their potential for miniaturization and flexibility \cite{BenitoPena.2016, Chatzidrosos.2021}.
A prominent example are magnetic-field sensors based on nitrogen-vacancy (NV) centers \cite{Johansson.2025}, where typically green excitation light ($\lambda_\mathrm{exc} \approx \SI{520}{\nano\meter}$) is delivered through the fiber to the NV-doped diamond, and red fluorescence (between $\SI{600}{\nano\meter}$ and $\SI{800}{\nano\meter}$) from the NV center is collected through the same fiber \cite{Graham.2023}.
In these sensors, the diamond, which contains multiple sensing NV centers, is positioned on top of the fiber or suspended within the fiber core \cite{Filipkowski.2022, Zhang.2022}.
The signal that inherits information about the magnetic-field amplitude and orientation is then optically read out by measuring the NV fluorescence.

However, in such fiber-based sensors, background light that overlaps spectrally with the fluorescence of the NV center reduces the signal contrast.
This background light is generated within the fiber and arises from Raman scattering and defect-related autofluorescence \cite{Udovich.2008}.
The light from inelastic Raman scattering originates from the interaction of the excitation light with phonon modes of fused silica, a material commonly used in optical fibers \cite{Bergler.2020, Dracinsky.2011, Plotnichenko.2000}, and shifts spectrally with the excitation wavelength.
In contrast, the fluorescence originating from several defects has a fixed emission wavelength but varies in intensity with the excitation light power \cite{Skuja.1998, Qian.2020, LoPiccolo.2021, Girard.2019}.
A critical defect center in terms of fluorescence for fiber-based applications with NV centers is the non-bridging oxygen hole center (NBOHC), which shows a broad emission band with a maximum at $\SI{1.92}{\electronvolt} = \SI{645}{\nano\meter}$ \cite{Sigel.1981, Vaccaro.2008b, Li.2019, Cannas.2006, Pacchioni.2000, Vaccaro.2008, Skuja.2012}.

\begin{figure*}[htbp]
    \centering
    \includegraphics[width=1\linewidth]{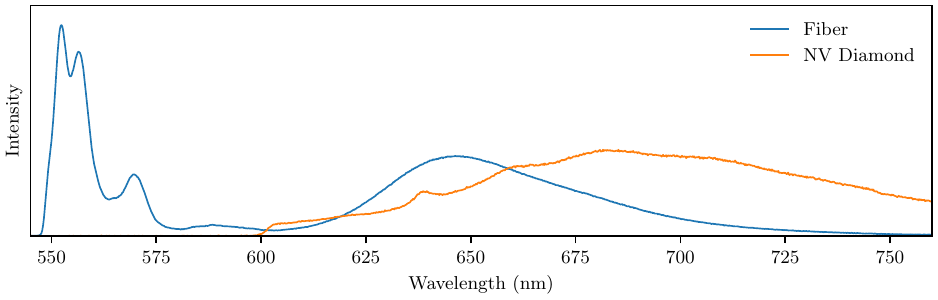}
    \caption{Typical fluorescence spectrum (orange) of a diamond containing multiple NV centers longpass-filtered at $\SI{600}{\nano\meter}$ (FELH0600) and measured spectrum (blue) of an FG050LGA optical fiber during transmission of $\SI{520}{\nano\meter}$ laser light longpass-filtered at $\SI{550}{\nano\meter}$ (FELH0550). 
    Both graphs are individually scaled to emphasize the overlap between the two spectra.
    The exact scaling of the fiber and NV-diamond spectra also varies with fiber length, number of NV centers, and diamond–fiber coupling efficiency.}
    \label{fig:NVdiamond_fiber}
\end{figure*}

While reflected and scattered excitation light can often be efficiently suppressed using spectral filters, fluorescence generated within the fiber itself, often called fiber autofluorescence or fiber luminescence \cite{Laemmel.2004, Li.2023}, can spectrally overlap with the NV emission and reduce the signal-to-noise ratio (SNR). 
This is illustrated in Fig.~\ref{fig:NVdiamond_fiber}, which shows the measured fiber background spectrum and the fluorescence spectrum of an NV-doped diamond.
As the NV fluorescence decreases, either due to fewer NV centers or a smaller diamond with constant NV doping, the constant fiber background limits the SNR.
Although amplitude modulation techniques \cite{Grace.2021}, spectral subtraction \cite{Ehrlich.2020}, bleaching \cite{Patrick.1994}, and time-resolved detection can mitigate some limitations, they do not resolve cases in which changes in excitation intensity directly affect the spectral shape and increase the experimental complexity \cite{Wang.2007, Lyu.2024, Xiang.2025}. 
For example, in experiments using NV centers, the relative population of $\mathrm{NV}^{-}$ and $\mathrm{NV}^{0}$, and thus the shape of the optical spectrum, changes with excitation light intensity \cite{CardosoBarbosa.2023}.
Therefore, modulating the excitation light intensity can affect the charge states and thus impact spectral measurements.
Other approaches to reduce background light, such as hollow-core fibers, which can have a lower background than solid-core fibers \cite{Yerolatsitis.2019}, are expensive and cannot easily be integrated into sensor setups that require a closed and solid fiber facet.
Therefore, they are not considered in this specific study.
Furthermore, the impact of practically relevant physical influences (cutting, temperature, bending, and scratching) on the spectral characteristics of commonly used and available optical fibers is rarely discussed, especially in the context of sensing with NV centers. 
Often, spectroscopic attributes of the amorphous silica or fused silica (\ce{SiO2}) of an optical fiber are analyzed.
However, the influence of the materials used in manufacturing, e.g., glues, tubing, ceramic ferrules, and fiber length, on the measured spectra, as well as the expected deviations between different optical fibers with fused-silica cores, remains unclear.

In our study, we analyze multiple commonly available optical fibers often used to guide light in the visible spectrum, especially in the field of NV centers.
We compare their obtained fluorescence spectra when coupling and guiding light of wavelength $\lambda_\mathrm{exc} \approx \SI{520}{\nano\meter}$ through the fiber, which is typically used to excite NV centers.
To identify the sources of the measured fiber background light, we investigate common materials used for fiber fabrication, e.g., connectors, tubing, glue, and different excitation wavelengths between $\SI{515}{\nano\meter}$ and $\SI{637}{\nano\meter}$.
Although typical excitation wavelengths of NV centers range from $\SI{515}{\nano\meter}$ to $\SI{532}{\nano\meter}$, spin-to-charge conversion sequences with NV centers require additional excitation wavelengths ($\lambda_\mathrm{exc} \approx \SI{594}{\nano\meter}$ and $\SI{637}{\nano\meter}$) \cite{Jaskula.2019}.
Therefore, we include these excitation wavelengths in our study.
We further analyze the fiber fluorescence as a function of the physical length, bending radius, coupling efficiency, and temperature. 
Finally, we report the optical spectra for several standard optical fibers. 
Based on these measurements, we provide a path to select the best optical fiber in terms of its autofluorescence properties and identify which materials and external influences most strongly impact the optical spectra.

\section{Background}
\begin{figure}[tbp]
    \centering
    \includegraphics[width=0.85\linewidth]{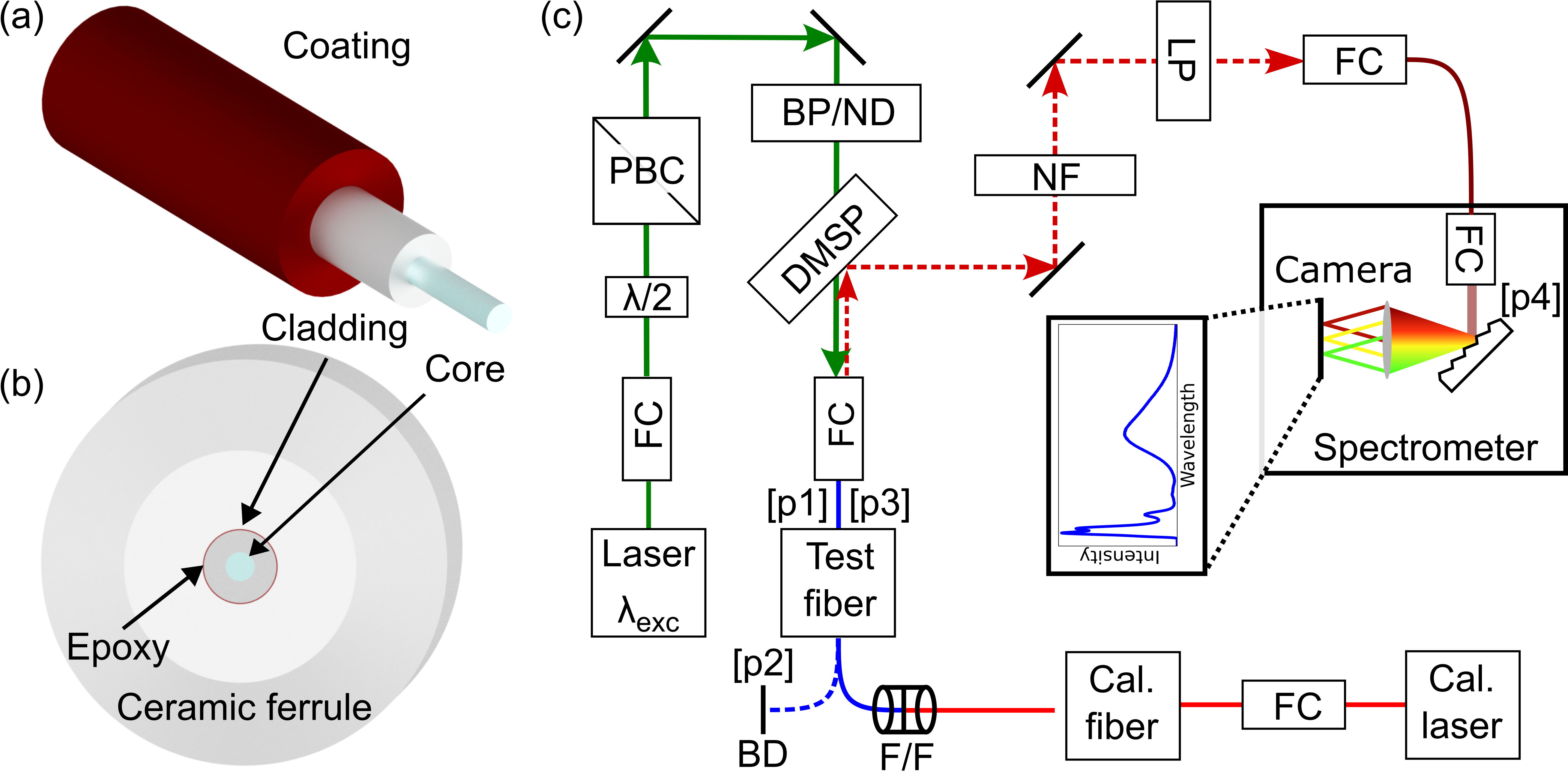}
    \caption{
    Common elements of a bare optical fiber and assembly in a fiber optic connector. 
    (a) Elements of a bare optical fiber, consisting of a core (light blue), cladding (gray), and protective coating (red).
    (b) Assembly of an optical fiber inside a ceramic fiber connector ferrule. 
    (c) Optical setup and measurement positions. 
    The optical setup consists of essentially two beam paths: Firstly, a fiber-coupled laser source (Laser $\lambda_\mathrm{exc}$), a fiber coupler (FC), a $\lambda/2$-plate, a polarizing beam cube (PBC), a bandpass (BP) filter, an optional neutral density (ND) filter, a dichroic mirror shortpass (DMSP), and another FC for coupling light in and out of the fiber under investigation (Test fiber). 
    The power of the excitation laser is adjusted using the optional ND filter and laser current, and it is fine-tuned with the $\lambda/2$-plate and PBC.
    Secondly, it consists of a beam path for measuring losses and the intrinsic light of the fiber between the test fiber and the spectrometer.
    For the analysis of losses, this path consists of a calibration laser (Cal. laser, CPS650F, Thorlabs), an FC with an attached calibration single-mode (SM) fiber (Cal. fiber, P1-460B-FC-2, Thorlabs), which then connects via a fiber-to-fiber connector (F/F) to the test fiber.
    Afterwards, the calibration light and intrinsic fiber light are transmitted through the test fiber and the connected FC, reflected at the DMSP, and, depending on the setup configuration, filtered from excitation light by a notch filter (NF) or a long-pass filter (LP). 
    Lastly, the remaining light is transmitted and focused onto the camera of the spectrometer via two fiber couplers, a blazed grating, and a lens.
    The marked positions p1-p4 indicate the measured positions used to determine losses, primarily due to coupling efficiencies.
    }
    \label{fig:faseraufbau}
    \label{fig:Setup_Exp}
\end{figure}

Optical fibers consist of multiple layers of materials with different purposes and properties. 
Typically, they have a light-guiding core surrounded by a cladding with a lower refractive index and a protective coating (polymer or metal), as shown in Fig.~\ref{fig:faseraufbau}~(a).
The optical properties of fibers are then specified by the numerical aperture (NA), which stems from the difference in refractive indices between the core and cladding materials, the core diameter, and the core material, which usually defines the transmission attenuation.
A typical fiber assembly consists of three major elements.
These are the optical fiber itself, which is further protected along its length by an additional plastic or metallic protective tubing, and the connector, in which the fiber is inserted and fixed with epoxy glue, as depicted in Fig.~\ref{fig:faseraufbau}~(b).

In the field of fiber-based NV sensors, which emit fluorescence light between $\SI{600}{\nano\meter}$ and $\SI{800}{\nano\meter}$ \cite{CardosoBarbosa.2023}, fused-silica-core fibers are an ideal candidate.
Low-\ce{OH} doped fused-silica fibers are especially suited, due to their low light damping in the visible spectrum and the absence of additional \ce{OH}-related attenuation peaks present in high-\ce{OH} doped fused-silica fibers~\cite{Humbach.1996}.
However, despite the suitable transmission properties of the material, it still shows intrinsic fluorescence and inelastic scattered light overlapping the NV centers' fluorescence, as shown in Fig.~\ref{fig:NVdiamond_fiber}.
To relate the background light intensity, especially the fiber autofluorescence, to the fluorescence intensity of a diamond containing NV centers, we compare autofluorescence from an optical fiber of $\SI{2}{\meter}$ length with the fluorescence of the same fiber when a diamond containing many NV centers is attached, similar to~\cite{Johansson.2025}.
In this comparison, the NV-doped diamond was chosen to be significantly brighter than the fiber background light, which is fulfilled for a $\SI{15}{\micro\meter}$-sized diamond with a NV doping of $\SI{3.5}{}\mathrm{ppm}$.
For a given fiber length, the autofluorescence generated within the optical fiber at a set excitation light intensity remains constant, while the fluorescence of the NV-doped diamond increases linearly with its volume.
To estimate the limit, when the autofluorescence becomes brighter than the NV-doped diamond, we divide the intensity of the NV-doped diamond with a diameter of $\SI{15}{\micro\meter}$ in the spectral range above $\SI{650}{\nano\meter}$ by the total intensity of the autofluorescence in the same spectral region.
From these results, an equal level of light intensity of fiber autofluorescence and NV fluorescence is reached for a $\SI{740}{\nano\meter}$-sized diamond, assuming a constant NV doping of $\SI{3.5}{}\,\mathrm{ppm}$.
Although this calculation directly applies only to fiber-based NV sensors, other sensors that rely on fluorescence measurements in a comparable spectral range could be affected similarly. 

Spectral features measured in optical fiber spectra that contribute to the fiber background light can be divided into two categories: wavelength-shifting and wavelength-static luminescence or fluorescence effects.
Wavelength-shifting effects, which are well known in fused silica, originate from inelastic scattering between light and optical or acoustical phonon modes \cite{Udovich.2008, Ippen.1972}.
These inelastic scattering effects, known as Brillouin and Raman scattering, exhibit a frequency shift independent of the excitation wavelength.
In the case of Brillouin scattering, typically a frequency shift of multiple $\SI{}{\giga\hertz}$ in fused-silica glass occurs due to the interaction between excitation light and acoustic phonon modes or density waves \cite{Tanaka.2017, Ippen.1972}.
In contrast, for the Raman-related Stokes or anti-Stokes shifts, which originate from inelastic scattering between excitation light and optical phonons, a frequency shift between excitation wavelength and Raman-shifted light of tens of $\SI{}{\tera\hertz}$ or hundreds of $\SI{}{\centi\meter^{-1}}$ is observed for fused silica~\cite{Dracinsky.2011, Udovich.2008}.
Therefore, for nitrogen-vacancy centers, where light is usually filtered below $\SI{650}{\nano\meter}$ and excitation light lies below $\lambda_\mathrm{exc} = \SI{637}{\nano\meter}$, Brillouin scattering has no noticeable impact, since wavelengths with a shift of much less than $\SI{1}{\nano\meter}$ relative to the excitation wavelength are filtered out.
In contrast, larger Raman-induced frequency shifts based on known Raman peaks with wavenumbers of up to $\SI{1198}{\centi\meter^{-1}}$ are to be expected in fused-silica glass \cite{Dracinsky.2011}.
This results in a spectral maximum with a shift of $\Delta\lambda_\mathrm{Raman} \approx \SI{34.5}{\nano\meter}$, for an excitation wavelength of $\lambda_\mathrm{exc} = \SI{520}{\nano\meter}$, while we also observe Raman-related peaks with shifts of $\Delta\lambda_\mathrm{Raman} \approx \SI{48}{\nano\meter}$ for all fibers.
Thus, we focus our analysis on Raman-related scattering effects.

In contrast to inelastic scattering, light-generating luminescent effects are known to originate from photoactive defects within the fused-silica glass commonly used for optical fibers in the visible spectrum and can spectrally overlap the NV centers' fluorescence.
In fused-silica glass, many defects are well known and identified~\cite{LoPiccolo.2021}. 
Most commonly known are NBOHCs, which have multiple absorption and emission bands between $\SI{1.9}{\electronvolt}$ and $\SI{6.8}{\electronvolt}$ \cite{Vaccaro.2008b, Hosono.2002}.
While one of these bands emits light with a maximum at $\lambda_\mathrm{emission}\approx \SI{645}{\nano\meter}$, which also aligns with our findings, the higher-energy bands are not detected in the spectral range between $\SI{550}{\nano\meter}$ and $\SI{800}{\nano\meter}$.
Other defects known in fused silica include, for example, ODC (oxygen-deficiency-related centers) and POR (peroxyl radical) \cite{Skuja.1998, LoPiccolo.2021, Pacchioni.2000}.
However, they exhibit different spectral features and are excited in different spectral ranges, which are not relevant for NV centers.
Thus, their impact is limited in terms of intensity and spectral overlap in this analysis \cite{Girard.2019}.
Although these defects have been extensively investigated in bulk glass materials, only a few studies focus on their influence in optical fibers for sensing applications, including practically relevant influences, and comparisons between different fiber types, which we addressed in this work.
This aims to provide a guide to common spectral defects observed in available fibers and their expected spectral features.

\section{Methods and Experimental Setup}
Before analyzing the measured light spectra of different fibers and materials, we discuss the optical setup and methods that enable comparable results.
The optical setup for the measurements of the spectra of all optical fibers is depicted in Fig.~\ref{fig:Setup_Exp}~(c).
Since the influence of different excitation lasers was investigated, different optical filters to block laser light at the spectrometer and filters to narrow the diode laser spectrum have been used for some lasers, as listed in Table~\ref{tab:Filter_combinations}.

\begin{table}[htbp]
\small
\setlength{\tabcolsep}{5pt}
  \centering
  \caption{Overview of the used lasers and corresponding varying optical components in the optical setup depending on the excitation wavelength.}
  \begin{tabular}{l c c c c}
    \toprule
    \textbf{Laser $\lambda_\mathrm{exc}$} & \textbf{Bandpass} & \textbf{DMSP} & \textbf{Longpass} & \textbf{Notch filter} \\
    \midrule
    $\SI{515.8}{\nano\meter}$ & FBH520-10 & 69-191 & FELH0550 & NF514-17 \\
    $\SI{520.2}{\nano\meter}$ & FBH520-10 & 69-191 & FELH0550 & --- \\
    $\SI{532.0}{\nano\meter}$ & --- & 69-191 & FELH0550 & NF533-17 \\
    $\SI{593.8}{\nano\meter}$ & --- & 69-193 & --- & NF594-23 \\
    $\SI{636.0}{\nano\meter}$ & FLH635-10 & 69-193 & FELH0650 & --- \\
    \bottomrule
  \end{tabular}
  
  \label{tab:Filter_combinations}
\end{table}

Throughout all measurements, fiber coupling was optimized after every exchange of the test fiber by pre-coupling the excitation laser light to a single-mode fiber and fine-tuning the alignment of the subsequently connected test fiber. 
The optical excitation power through the test fiber for comparative measurements was maintained constant, and the laser power was measured at position [p2].
During the measurements of the optical spectra, the excitation light guided through the test fiber was directed to a beam dump (BD), and light generated inside the fiber was guided and transmitted in the reverse direction to the spectrometer as depicted in Fig.~\ref{fig:Setup_Exp}~(c).
To rule out any influence of the black-anodized aluminum of the BD on the measured spectra, the light transmitted through the fiber was directed onto different beam-dump materials, none of which showed a measurable change.

Furthermore, since optical fibers with different numerical apertures, materials, and core diameters were investigated without changing fiber couplers, losses due to different coupling efficiencies had to be measured to enable the comparison of spectral features and their intensities.
These losses between the test fiber and the spectrometer were calibrated by connecting a single-mode fiber-coupled laser (Cal.\ laser and Cal.\ fiber) to each test fiber using the (F/F) connector and measuring the transmitted power at the spectrometer [p4]. 
We used a calibration laser with a wavelength of $\lambda_\mathrm{cal}=\SI{650}{\nano\meter}$, whose power was measured at [p3] and in front of the grating [p4] to calculate coupling losses.
The spectrometer consists of a blazed grating, whose reflected light is then focused on a camera sensor. 
Wavelength calibration of the spectrometer was performed by sending multiple laser light sources with known wavelengths through the connected optical fiber, mapping the corresponding pixel positions to these wavelengths, and interpolating between them using a 2nd-degree polynomial.
Since the transmission curves of the optical elements are partially unknown, some deviation between the exact and measured intensities is expected.
However, we rely on relative changes that can be directly compared.
All combinations of optical fiber configurations are listed in Table~\ref{tab:Fiber_combinations}.

\begin{table*}[htbp]
\small
\setlength{\tabcolsep}{2.8pt}
{
\renewcommand{\arraystretch}{1.6}
\caption{Overview of fiber assemblies. Fiber names, dimensions, numerical aperture (NA) values, and material naming refer to the original naming and data sheets of Thorlabs. In the column 'Assembly', FC/PC fibers that have been fully assembled and connectorized by the manufacturer are marked, e.g., as 'Thorlabs', while self-assembled fibers are marked as 'Own'. For our own fibers, black tubing (FT020-BK) and black fiber connector boots from Thorlabs were used. The compared commercially assembled fibers have a tubing with a diameter of $\SI{3}{\milli\meter}$ and with a yellow or orange color. All listed fibers, besides No.\,13 and 14, have a length of $\SI{2}{\meter}$, while No.\,13 and 14 have a length of $\SI{5}{\meter}$. The type of fiber, in terms of single-mode (SM) and multi-mode (MM), is given in the column 'Modes'.}
\begin{tabular}{l|ccccccc}
\toprule
\textbf{No., Fiber name} & \textbf{\diameter\, Core} & \textbf{\diameter\, Cladding} & \textbf{\diameter\,  Coating} & \textbf{NA} & \textbf{Epoxy} & \textbf{Modes} & \textbf{Assembly} \\
\midrule
1, SM450 & $\num{3.45}^{+0.65}_{-0.65}\, \SI{}{\micro\meter}$ & $\num{125}^{+1}_{-1}\, \SI{}{\micro\meter}$ & $\num{245}^{+15}_{-15}\, \SI{}{\micro\meter}$ & $\SI{0.13}{}$ & unknown & SM & Thorlabs \\ 
2, SM600 & $\num{4.45}^{+0.85}_{-0.85}\, \SI{}{\micro\meter}$ & $\num{125}^{+1}_{-1}\, \SI{}{\micro\meter}$ & $\num{245}^{+15}_{-15}\, \SI{}{\micro\meter}$ & $\SI{0.12 \pm 0.02}{}$ & 353NDPK & SM & Own \\
3, FG010DA & $\num{10}^{+3}_{-3}\, \SI{}{\micro\meter}$ & $\num{125}^{+2}_{-2}\, \SI{}{\micro\meter}$ & $\num{245}^{+10}_{-10}\, \SI{}{\micro\meter}$ & $\SI{0.1 \pm 0.015}{}$ & unknown & MM & Thorlabs \\
4, FG050LGA & $\num{50}^{+1}_{-1}\, \SI{}{\micro\meter}$ & $\num{125}^{+1}_{-2}\, \SI{}{\micro\meter}$ & $\num{250}^{+10}_{-10}\, \SI{}{\micro\meter}$ & $\SI{0.22 \pm 0.02}{}$ & unknown & MM & Thorlabs \\ 
5, FG050LGA & $\num{50}^{+1}_{-1}\, \SI{}{\micro\meter}$ & $\num{125}^{+1}_{-2}\, \SI{}{\micro\meter}$ & $\num{250}^{+10}_{-10}\, \SI{}{\micro\meter}$ & $\SI{0.22 \pm 0.02}{}$ & 353NDPK & MM & Own \\ 
6, FG050LGA & $\num{50}^{+1}_{-1}\, \SI{}{\micro\meter}$ & $\num{125}^{+1}_{-2}\, \SI{}{\micro\meter}$ & $\num{250}^{+10}_{-10}\, \SI{}{\micro\meter}$ & $\SI{0.22 \pm 0.02}{}$ & F112 & MM & Own \\ 
7, FG050LGA & $\num{50}^{+1}_{-1}\, \SI{}{\micro\meter}$ & $\num{125}^{+1}_{-2}\, \SI{}{\micro\meter}$ & $\num{250}^{+10}_{-10}\, \SI{}{\micro\meter}$ & $\SI{0.22 \pm 0.02}{}$ & R\&G HT2 & MM & Own \\
8, FG050UGA & $\num{50}^{+1}_{-1}\, \SI{}{\micro\meter}$ & $\num{125}^{+1}_{-2}\, \SI{}{\micro\meter}$ & $\num{250}^{+10}_{-10}\, \SI{}{\micro\meter}$ & $\SI{0.22 \pm 0.02}{}$ & 353NDPK & MM & Own \\ 
9, FG0105LCA & $\num{105}^{+1}_{-3}\,\SI{}{\micro\meter}$ & $\num{125}^{+1}_{-2}\, \SI{}{\micro\meter}$ & $\num{250}^{+10}_{-10}\, \SI{}{\micro\meter}$ & $\SI{0.22 \pm 0.02}{}$ & 353NDPK & MM & Own \\ 
10, UM22-100 & $\num{100}^{+3}_{-3}\,\SI{}{\micro\meter}$ & $\num{110}^{+3}_{-3}\, \SI{}{\micro\meter}$ & $\num{124}^{+3}_{-3}\, \SI{}{\micro\meter}$ & $\SI{0.22 \pm 0.02}{}$ & 353NDPK & MM & Own \\ 
11, FP200ERT & $\num{200}^{+5}_{-5}\, \SI{}{\micro\meter}$ & $\num{225}^{+5}_{-5}\, \SI{}{\micro\meter}$ & $\num{500}^{+30}_{-30}\, \SI{}{\micro\meter}$ & $\SI{0.50}{}$ & 353NDPK & MM & Own \\
12, AFM200L & $\num{200}^{+3}_{-3}\, \SI{}{\micro\meter}$ & $\num{220}^{+3.3}_{-3.3}\, \SI{}{\micro\meter}$ & $\num{300}^{+30}_{-30}\, \SI{}{\micro\meter}$ & $\SI{0.22}{}$ & 353NDPK & MM & Own \\ 
13, FT400EMT & $\num{400}^{+8}_{-8}\, \SI{}{\micro\meter}$ & $\num{425}^{+10}_{-10}\, \SI{}{\micro\meter}$ & $\num{730}^{+30}_{-30}\, \SI{}{\micro\meter}$ & $\SI{0.39}{}$ & unknown & MM & Thorlabs \\
14, OM2 & $\num{50}^{+2.5}_{-2.5}\, \SI{}{\micro\meter}$ & $\num{125}^{+1}_{-1}\, \SI{}{\micro\meter}$ & $\num{242}^{+5}_{-5}\, \SI{}{\micro\meter}$ & $\SI{0.2}{}$ & unknown & MM & Corning \\
\bottomrule
\end{tabular}
\label{tab:Fiber_combinations}
}
\end{table*}

\section{Results}
To identify the origin of the observed background light from an optical fiber, we investigate the materials used in the assembly, physical influences, and its spectral characteristics.

\subsection{Materials for Assembly}
\begin{figure}[htbp]
    \centering
    \includegraphics[width=1\columnwidth]{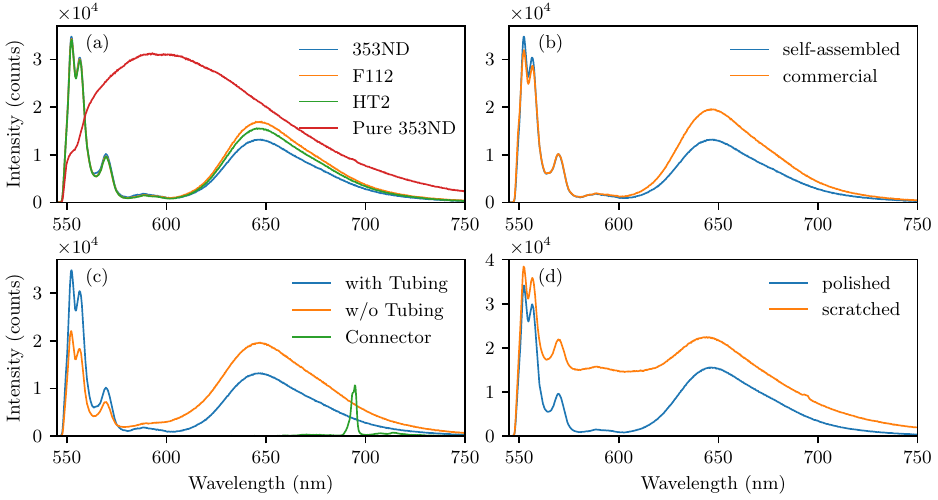}
    \caption{
    Impact on the optical spectra of the used materials for a fiber assembly. 
    (a) Measured optical spectra of the same optical fiber using different epoxy glues (Table~\ref{tab:Fiber_combinations}, Nos.\,5-7) and a connector fully filled with Epo-tek 353ND without any fiber inserted. 
    (b) Optical spectra of a commercially fully assembled fiber (Table~\ref{tab:Fiber_combinations}, No.\,4) and a self-assembled fiber (No.\,5). 
    (c) Optical spectrum of an empty connector in comparison to the spectrum of the fiber in Table~\ref{tab:Fiber_combinations}, No.\,5, with and without a furcation tubing. 
    (d) Spectra before and after scratching a fully assembled fiber tip (Table~\ref{tab:Fiber_combinations}, No.\,7).
    }
    \label{fig:epoxy_conn_tub_poly}
\end{figure}

We first analyze materials used for the assembly of the FC/PC-connectorized and tubed fiber.
For this, different epoxy glues (Epo-tek 353ND, Loctite Eccobond F112, R\&G HT2) that secure the optical fiber in the FC/PC connector were investigated and also compared with a fully assembled fiber from a commercial manufacturer (Thorlabs). 
Respective results are depicted in Fig.~\ref{fig:epoxy_conn_tub_poly}~(a) and (b).
Furthermore, in Fig.~\ref{fig:epoxy_conn_tub_poly}~(c), the difference between a fully assembled fiber and the same type of fiber with a single connector without a protective tubing was measured. 
In addition, the optical spectrum of an empty connector with a ceramic ferrule (30128C3, Thorlabs) without any inserted fiber was measured and spectrally compared.
Moreover, the importance of the polishing process is investigated in Fig.~\ref{fig:epoxy_conn_tub_poly}~(d) that shows the change in the spectrum after scratching a fiber's facet.

The shapes of the spectra in Fig.~\ref{fig:epoxy_conn_tub_poly}~(a) are similar in their behavior and show two regions of higher spectral intensity below $\SI{600}{\nano\meter}$ and above $\SI{600}{\nano\meter}$, yet the intensity of the spectral feature around $\SI{645}{\nano\meter}$ changes slightly. 
In contrast, the measured optical spectrum in Fig.~\ref{fig:epoxy_conn_tub_poly}~(a) of a connector filled with pure epoxy 353ND shows an entirely different shape. 
Since the optical spectra of the fibers practically overlap between $\SI{550}{\nano\meter}$ and $\SI{600}{\nano\meter}$, where 353ND has its maximum intensity, this indicates that the influence of the epoxy is not significant.
A possible reason for the different autofluorescence intensities between the same fibers with different epoxies above $\SI{600}{\nano\meter}$ could be a slight difference in the fiber material or the influence of heating.
In the case of the Epo-tek 353ND, the whole fiber had to be heated to cure the epoxy \cite{Witcher.2013}.
Furthermore, we also conclude that the connector does not have a significant influence on the spectrum, as long as the ceramic ferrule is not directly illuminated, which will be analyzed in more detail in Fig.~\ref{fig:length_bend_coupling}~(d).
Also, the observed spectral features below $\SI{600}{\nano\meter}$ and above $\SI{600}{\nano\meter}$ will be analyzed further in Fig.~\ref{fig:power_scaling} and Fig.~\ref{fig:raman_wavenumber_wavelength}, yet they already indicate that the spectrum inherits at least two origins: the Raman-related scattered light and fluorescence-related effects in the range of $\SI{600}{\nano\meter}$ to $\SI{750}{\nano\meter}$.
In Fig.~\ref{fig:epoxy_conn_tub_poly}~(b), a commercial fiber and a self-assembled fiber are compared.
Similarly to Fig.~\ref{fig:epoxy_conn_tub_poly}~(a), it shows practically no difference in the range between $\SI{550}{\nano\meter}$ and $\SI{600}{\nano\meter}$, yet a significant difference above $\SI{600}{\nano\meter}$.
To exclude the influence of protective tubing material in the observed spectra, we analyze the spectra of two identical optical fibers, of which only one fiber is fully assembled, and another fiber is left unprotected and only connectorized on one side, in Fig.~\ref{fig:epoxy_conn_tub_poly}~(c).
Similarly, we measure the optical spectrum of a bare ceramic-ferrule connector to analyze its influence on the optical spectrum. 
The spectrum of the optical fiber without a tubing shows a higher intensity in the spectral region above $\SI{600}{\nano\meter}$ compared to the fiber with a tubing and a second connector attached. 
In the spectral range below $\SI{600}{\nano\meter}$, the behavior is the opposite. 
We attribute this difference to the unpolished end, slightly lower coupling efficiencies, and the external coupling of light along the entire length of the fiber.
However, adding protective tubing has no negative impact on the spectral features above $\SI{600}{\nano\meter}$.
As observed in Fig.~\ref{fig:epoxy_conn_tub_poly}~(c), the spectral behavior of a ceramic connector illuminated with $\SI{520}{\nano\meter}$ excitation light shows a distinct maximum at $\approx \SI{695}{\nano\meter}$, which is not visible when illuminating through a partially or fully connectorized multi-mode optical fiber that is properly fiber-coupled.
In Fig.~\ref{fig:epoxy_conn_tub_poly}~(d), changes caused by improper polishing and scratching of the fiber facet are investigated. 
Here, the spectra were measured after the fiber facet was polished properly and after scratching the polished facet using a coarse $\SI{6}{\micro\meter}$-grid size diamond polishing pad (LF6D, Thorlabs).
The observed spectra of the scratched fiber show increased intensity across the entire spectrum, possibly due to more back-reflections of the laser light and contamination at the scratched surface.
Additionally, a local maximum at $\SI{695}{\nano\meter}$ indicates that material from the connector accumulated on the fiber facet.
The observations in Fig.~\ref{fig:epoxy_conn_tub_poly}~(d) thereby identify a careful polishing process as an essential step to minimize the fiber background light.
In conclusion, the fiber assembled with Epo-Tek 353ND provides the lowest amount of autofluorescence. 
Furthermore, our self-assembled fibers provide comparable spectra to the same commercially connectorized optical fiber.
Subsequently, in the following measurements, the self-assembled fibers were joined with Epo-tek 353ND and a suitable connector with a ceramic ferrule.

\subsection{Influence of External Factors}

\begin{figure}[htbp]
    \centering
    \includegraphics[width=1\columnwidth]{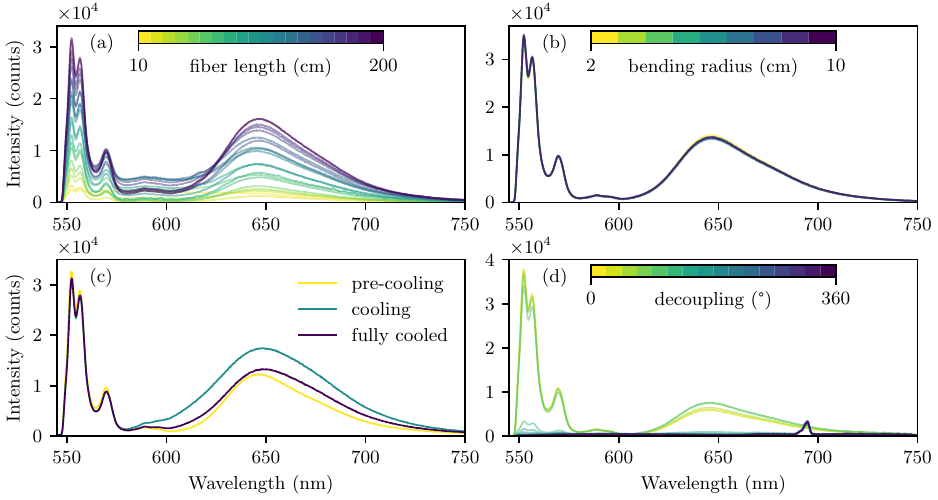}
    \caption{
    Influence of external factors on the fiber background spectrum.
    (a) Detected spectra after cutting the length of the fiber stepwise. The fiber was equipped with a connector only on one side, and the protective tubing was shortened together with the optical fiber.
    (b) Detected spectra when bending the fiber around a cylinder with varying diameter.
    (c) Influence of rapid cooling on the received light spectrum. Here, a large portion of the fiber was placed in a container filled with liquid nitrogen. The orange cooling spectrum was taken a few seconds after inserting it in the liquid nitrogen, while the green fully cooled spectrum was measured after thermalization.
    (d) Influence of misalignment of the fiber coupling. The unit $\SI{}{\degree}$ represents the turns in $\SI{}{\degree}$ on a mirror before the fiber, while $\SI{0}{\degree}$ represents pre-aligned position. Above a rotation of $\approx\SI{200}{\degree}$, the measured spectra remain constant and no more measurable light is guided. 
    }
    \label{fig:length_bend_coupling}
\end{figure}

Besides the choice of materials used for the assembly of the optical fibers, other influences potentially impacting the spectra are analyzed, such as imperfect fiber coupling, stress-induced effects, and fiber length.
The change of the spectrum is investigated in terms of fiber length, physical bending, and rapid cooling of the fiber in Fig.~ \ref{fig:length_bend_coupling}~(a)-(c). 
The influence of imperfect fiber coupling of the excitation laser to the test fiber is studied in Fig.~\ref{fig:length_bend_coupling}~(d).
For all spectra in Fig.~\ref{fig:length_bend_coupling}, the same fiber type (FG050LGA) and optical elements were used (Table~\ref{tab:Fiber_combinations}, No.\,5). 

In Fig.~\ref{fig:length_bend_coupling}~(a), the change in intensity for different lengths of the optical fiber is analyzed.
As the fiber length decreases, the overall intensity of the spectrum also decreases.
This observation indicates that the measured light is generated homogeneously along the whole length of the optical fiber.
Slight deviations in the decrease are observed, which we attribute to imperfect cutting of the fiber with a scissor or remaining dirt, leading to back reflections of the excitation light at the cut end of the fiber, similar to what is discussed in (e).
As depicted in (b), we find that bending of the fiber has no measurable effect on the fluorescence spectrum in the range larger than the minimal short-term bend radius ($r_\mathrm{min}=\SI{16}{\milli\meter}$).
In Fig.~\ref{fig:length_bend_coupling}~(c), the spectral change due to cooling is analyzed.
Here, approximately half of the fiber was positioned in a container, which was then filled with liquid nitrogen.
The spectra were measured before cooling (pre-cooling), a few seconds after insertion of the fiber into liquid nitrogen (cooling), and after thermalization (fully cooled) of the fiber in liquid nitrogen.
While the first part of the spectrum in Fig.~\ref{fig:length_bend_coupling}~(c) below $\SI{600}{\nano\meter}$ remains rather constant during cooling, a clear change in fluorescence properties can be observed in the range between $\SI{600}{\nano\meter}$ and $\SI{750}{\nano\meter}$.
First, fluorescence increases during cooling, then slightly decreases to a higher intensity than in the pre-cooled state.
This observation leads to the conclusion that the spectral features have a different origin and are differently impacted by thermal shocks.
The measured increase of fluorescence during the cooling could indicate that the fluorescence is partially stress-related, potentially due to different expansion coefficients of the acrylate coating and pure silica core.
The increase in fluorescence intensity compared to the pre-cooled state would fit an expected increase of NBOHC-related fluorescence with decreasing temperatures \cite{Vaccaro.2008}.
In Fig.~\ref{fig:length_bend_coupling}~(d), the influence of coupling precision is analyzed, revealing essentially two different spectra.
Either the light is coupled and guided through the fiber, resulting in a spectrum independent of the angle of the in-coupled light, or it is not guided, and the ceramic ferrule is illuminated, producing a spectrum similar to that observed for an empty connector in Fig.~\ref{fig:epoxy_conn_tub_poly}~(c).
Furthermore, it can be observed that fluorescence increases slightly before light is no longer coupled, which could indicate that the cladding also contributes to the fluorescent signal, which was also found in other publications \cite{Bianco.2021}.
Thus, the influence of optimal coupling to the core of the multi-mode (MM) fiber seems uncritical as long as light is guided.

\subsection{Influence of Excitation Light}

\begin{figure}[htbp]
    \centering
    \includegraphics[width=1\linewidth]{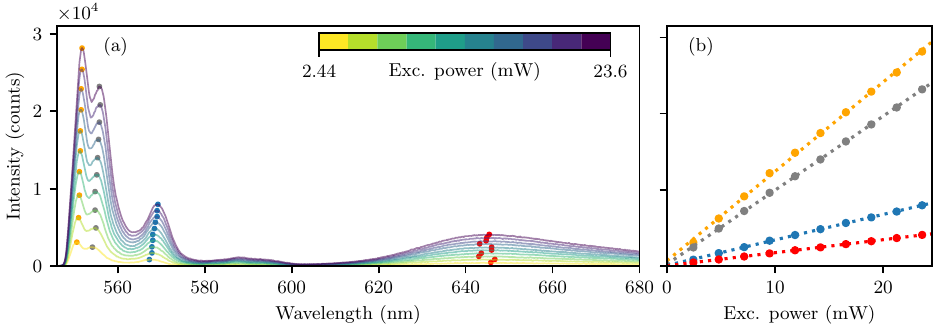}
    \caption{
    Measured optical spectra of a fiber (Table~\ref{tab:Fiber_combinations}, No.\,5) for varying laser power (a). 
    For each spectrum, local maxima are marked.
    In (b), the retrieved intensities of each local maximum are plotted against excitation power and fit with a linear regression.
    The colors of the markers in (b) correspond to the markers in (a).
    }
    \label{fig:power_scaling}
\end{figure}

To further understand the observed spectral features and their origins, as well as their scaling with changing excitation power, the measured optical spectra are analyzed as the laser light power increases in Fig.~\ref{fig:power_scaling}.

As depicted in Fig.~\ref{fig:power_scaling}~(a), the total spectral intensity increases with increasing excitation light power.
The spectral peak intensities in (a) scale linearly with excitation power, but each peak has a distinct slope, as shown in Fig.~\ref{fig:power_scaling}~(b).
Besides a change in intensity, the corresponding wavelength of each maximum also changes for the peaks in Fig.~\ref{fig:power_scaling}~(a) below $\SI{600}{\nano\meter}$.
The behavior of these shifting peaks indicates that the spectral peaks correspond to Raman-related shifts, caused by a wavelength shift of the utilized diode laser of about $\SI{2}{\nano\meter}$ rather than an intensity-related shift.
In contrast, the corresponding wavelength of the maximum in (a) above $\SI{600}{\nano\meter}$ does not show a clear trend in one direction. 
Instead, the position of the maximum fluctuates around a central value of $\lambda_\mathrm{NBOHC}=\SI{644.9}{\nano\meter}\pm\SI{1.3}{\nano\meter}$, matching an expected luminescence maximum of NBOHC defects \cite{Vaccaro.2008b}. 

\begin{figure}[htbp]
    \centering
    \includegraphics[width=1\linewidth]{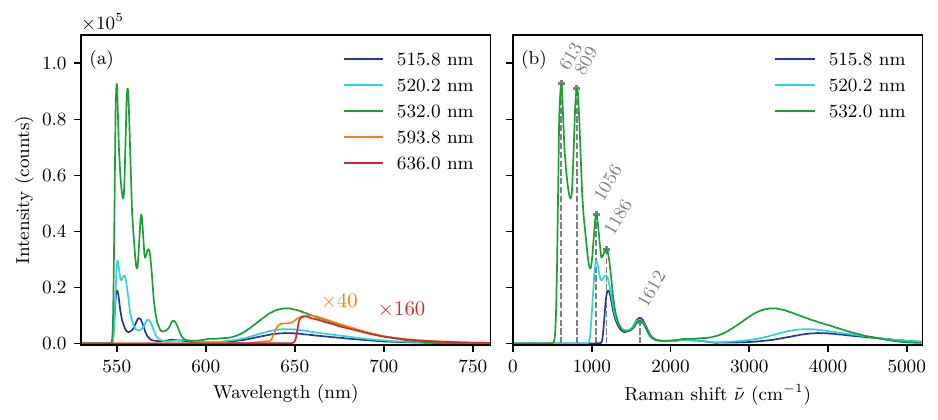}
   \caption{
    Directly measured optical spectra and calculated Raman-shifted spectra for five different excitation wavelengths.
    (a) Measured light intensity plotted against the wavelength of the same optical fiber (Table~\ref{tab:Fiber_combinations}, No.\,5) for different excitation wavelengths.
    The depicted optical spectra are filtered according to Table~\ref{tab:Filter_combinations}.
    (b) Intensities from the optical spectra in (a) plotted against the Raman-shifted wavenumbers for three excitation wavelengths ($\lambda_\mathrm{exc} = \SI{515}{\nano\meter}$, $\SI{520}{\nano\meter}$, $\SI{532}{\nano\meter}$). 
    }
    \label{fig:raman_wavenumber_wavelength}
\end{figure}

To analyze the shifts observed in Fig.~\ref{fig:power_scaling}~(a) further, we measure the optical spectra for five different excitation light wavelengths in Fig.~\ref{fig:raman_wavenumber_wavelength}~(a) and calculate the Raman shifted spectra for three wavelengths in Fig.~\ref{fig:raman_wavenumber_wavelength}~(b).
In these measurements, the optical spectra of fiber No.\,5 were filtered, according to Table~\ref{tab:Filter_combinations}, to block excitation laser light from reaching the detector.
Therefore, the optical spectra with excitation wavelengths $\lambda_\mathrm{exc}$ of $\SI{515}{\nano\meter}$, $\SI{520}{\nano\meter}$, and $\SI{532}{\nano\meter}$ were effectively filtered below $\SI{550}{\nano\meter}$, while the spectra at $\lambda_\mathrm{exc} = \SI{594}{\nano\meter}$ and $\SI{636}{\nano\meter}$ were filtered below $\SI{650}{\nano\meter}$.

As shown in Fig.~\ref{fig:raman_wavenumber_wavelength}~(a), the maxima in the optical spectra below $\SI{600}{\nano\meter}$ clearly shift with the excitation wavelengths, as also visible in Fig.~\ref{fig:power_scaling}~(a).
Furthermore, the corresponding maxima in this range remain nearly constant, but are damped by the transmission properties close to the cut-off wavelength of the longpass filters.
In contrast, the optical spectrum above $\SI{600}{\nano\meter}$ remains constant in its shape and maximum position, but the overall intensity changes significantly.
This change in intensity is especially evident for the curve obtained for $\lambda_\mathrm{exc}=\SI{594}{\nano\meter}$ and $\SI{636}{\nano\meter}$, which were divided by factors of $40$ and $160$, respectively, to overlap the curve measured for $\lambda_\mathrm{exc}=\SI{532}{\nano\meter}$.
The observed increase in fluorescence intensity of the measured peak between $\SI{600}{\nano\meter}$ and $\SI{750}{\nano\meter}$ matches the properties of an NBOHC defect, which has an absorption band at $\approx \SI{630}{\nano\meter}$ \cite{Vaccaro.2008b, Pacchioni.2000}. 
To verify for $\lambda_\mathrm{exc}=\SI{515}{\nano\meter}$, $\SI{520}{\nano\meter}$, and $\SI{532}{\nano\meter}$ that the observed shifts below $\SI{600}{\nano\meter}$ are related to an inelastic scattering process, the Raman-shifted spectra are plotted in Fig.~\ref{fig:raman_wavenumber_wavelength}~(b). 
There, the $x$-axis is rescaled to Raman-shifted wavenumbers $\tilde{\nu}$, which were calculated from the intensity values at each measured wavelength $\lambda_\mathrm{meas}$, according to

\begin{equation}
    \tilde{\nu} = \frac{1}{\lambda_\mathrm{exc}} - \frac{1}{\lambda_\mathrm{meas}} \,.
    \label{eq_wavenumbercalc}
\end{equation}

The excitation wavelength $\lambda_\mathrm{exc}$ used in eq. \ref{eq_wavenumbercalc} was determined by measuring the laser light with a spectrometer (Yokogawa AQ6373) at the optical power of $\SI{2}{\milli\watt}$ at the end of the tested fiber.
As depicted in the Raman-shifted spectra in Fig.~\ref{fig:raman_wavenumber_wavelength}~(b), all maxima positions besides the observed intensity maximum at $\SI{645}{\nano\meter}$ in Fig.~\ref{fig:raman_wavenumber_wavelength}~(a) overlap, which is especially visible for the maximum at $\SI{1612}{\per\centi\meter}$.
For the spectra with the excitation wavelengths $\lambda_\mathrm{exc}=\SI{515}{\nano\meter}$ and $\SI{520}{\nano\meter}$, not all the peaks visible for $\lambda_\mathrm{exc}=\SI{532}{\nano\meter}$ are observed because of the filters' cut-off wavelength relative to the excitation wavelength.
In total, five Raman peaks are observed at $\tilde{\nu} = \SI{613}{\per\centi\meter}$, $\SI{809}{\per\centi\meter}$, $\SI{1056}{\per\centi\meter}$, $\SI{1186}{\per\centi\meter}$, and $\SI{1612}{\per\centi\meter}$.
These peaks also exist in the spectra retrieved for $\lambda_\mathrm{exc}=\SI{594}{\nano\meter}$ and $\lambda_\mathrm{exc}=\SI{636}{\nano\meter}$.
However, they overlap with the high-intensity NBOHC fluorescence spectrum and are two orders of magnitude less intense.
Thus, they would not be visible in Fig.~\ref{fig:raman_wavenumber_wavelength}~(b) and are, therefore, not plotted.
In conclusion, this fits the known Raman spectrum of fused-silica glass commonly used for optical fibers \cite{Yang.2023, Dracinsky.2011}, although the peak at $\tilde{\nu} = \SI{1612}{\per\centi\meter}$ is usually not reported for fused-silica glass, and Raman-related peaks below $\SI{600}{\per\centi\meter}$ can not be detected with the given optical filters.

\subsection{Comparison of Different Optical Fibers}

\begin{figure}[htbp]
    \centering
    \includegraphics[width=1\linewidth]{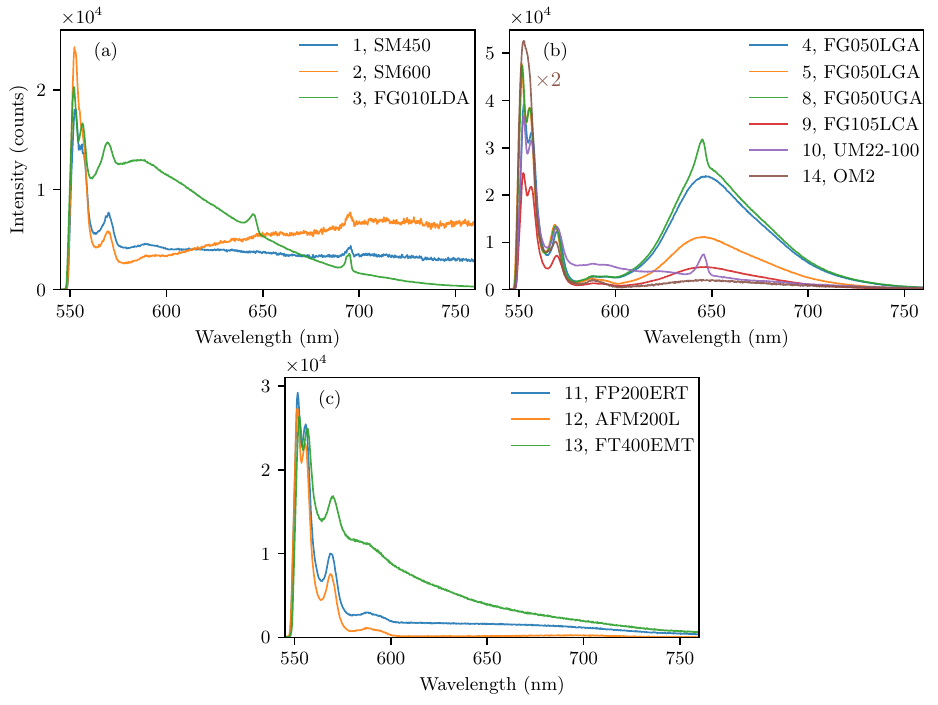}
    \caption{
    Measured spectra of different optical fibers as listed in Table~\ref{tab:Fiber_combinations}.
    The spectra are separated into three subplots for optical fibers with varying core sizes and numerical apertures.
    Plot (a) shows spectra of SM and MM fibers with small core sizes $<\SI{10}{\micro\meter}$ and numerical apertures of $\approx \SI{0.1}{}$.
    In (b), spectra of MM fibers with core sizes between $\SI{50}{\micro\meter}$ and $\SI{105}{\micro\meter}$ and numerical apertures of $\approx \SI{0.2}{}$ are depicted.
    The intensity of the spectrum of fiber No. 14. has been multiplied by a factor of $0.5$ for better visualization and is thus $2\times$ brighter than depicted. 
    In (c), optical spectra of MM fibers with comparatively large core diameters $>\SI{200}{\micro\meter}$ and numerical apertures between $\SI{0.22}{}$ and $\SI{0.5}{}$ are shown.
    }
    \label{fig:all_fibers}
\end{figure}

Lastly, the spectral shape and relative intensities of different spectrally suitable (VIS/NIR) optical fibers with varying core sizes and numerical apertures are measured and compared in Fig.~\ref{fig:all_fibers}. 
In Fig.~\ref{fig:all_fibers}~(a-c), the detected spectra of all listed fibers in Table~\ref{tab:Fiber_combinations}, without the duplicate fibers No.\,6 and 7, are shown sorted by fiber core diameter and numerical aperture. 
As depicted in Fig.~\ref{fig:all_fibers}~(a), the spectra of all fibers with core sizes below $\SI{10}{\micro\meter}$ and numerical apertures of $\approx \SI{0.1}{}$ show the Raman peaks corresponding to the fused-silica glass in addition to a peak at $\approx \SI{695}{\nano\meter}$, which is also visible in the spectrum of an empty connector in Fig.~\ref{fig:epoxy_conn_tub_poly}~(c).
Additionally, the multi-mode fiber (FG010LDA) shows a further local maximum at $\SI{645}{\nano\meter}$ and generally exhibits higher intensity values below $\SI{650}{\nano\meter}$, which decrease with increasing wavelength. 

In Fig.~\ref{fig:all_fibers}~(b), all fibers show the same distinctive Raman peaks below $\SI{600}{\nano\meter}$.
However, their intensities differ slightly, especially for fiber OM2, which has been multiplied by a factor of $0.5$ for better comparison in Fig.~\ref{fig:all_fibers}~(b) and still shows the highest Raman intensity.
The intensity difference of the Raman peaks could be related to the Germanium doping and manufacturing process, as this is the only graded-index, Germanium-doped optical fiber compared to all other step-index fibers.
Additionally, all measured multi-mode fibers with core sizes between $\SI{50}{\micro\meter}$ and $\SI{105}{\micro\meter}$ show a broad fluorescence peak between $\SI{600}{\nano\meter}$ and $\SI{750}{\nano\meter}$, except for the UM22-100 fiber, which only shows a sharp peak at $\approx \SI{645}{\nano\meter}$.
Moreover, the fibers UM22-100 and FG050UGA show an additional sharp peak at $\approx\SI{645}{\nano\meter}$, similar to fiber FG010LDA.
We assume the additional sharp peak could be related to a hydroxyl-group-related Raman scattering effect, since its position also shifts with the excitation wavelength and the peak is only visible in the OH-enriched fiber FG050UGA compared to the otherwise similar low-\ce{OH} FG050LGA fiber \cite{Plotnichenko.2000}.
We suppose that the observed difference of the broad NBOHC-related fluorescence peak between the same type of fiber (No.\,5 and 4) in (b) is mainly related to different bleaching duration, different purity of the glass, and different heat treatment of the fused silica of the fiber \cite{Witcher.2013, Li.2019}. 
For the fibers that also show the broad fluorescence peak, we assume the difference could be related to a different purity or the drawing process. 
In conclusion, for the depicted fibers in (b) in the relevant spectral range $>\SI{600}{\nano\meter}$, no explicit dependency in terms of fluorescence versus core diameter can be drawn.

For larger core diameters $\geq\SI{200}{\micro\meter}$ as depicted in Fig.~\ref{fig:all_fibers}~(c), again Raman peaks below $\SI{600}{\nano\meter}$ are observed.
In comparison, the aluminium-coated fiber AFM200L provides the absolute lowest fluorescence intensity.
However, the fiber FP200ERT, which has an identical core diameter but a higher numerical aperture, also shows a low fluorescence intensity compared to all fibers in Fig.~\ref{fig:all_fibers}.
For the fibers that show an increased background below $\SI{650}{\nano\meter}$, which are unrelated to the Raman peaks, we find that these fibers (FG010LDA, FP200ERT, FT400EMT) use a polymer cladding, which might contribute more than a doped fused-silica cladding.

In summary, we find that many of the measured fibers exhibit emission spectra with distinct spectral behaviors due to luminescence and inelastic Raman scattering processes.
Additionally, our measurements indicate that the spectral characteristics are not directly correlated with the core diameter and numerical aperture. 
In total, for wavelengths above $\SI{650}{\nano\meter}$ in the relevant NV fluorescence spectrum, the fibers AFM200L, followed by UM22-100, exhibit the lowest fluorescence. 
However, the total intensity of the fiber FP200ERT is also low, and it has a higher NA and therefore collects light with a larger acceptance angle.
Also, when selecting the ideal fiber for a sensing application, several additional factors must be considered.
Firstly, the amount of light coupled to a fiber without any attached optics depends on the numerical aperture and the core diameter in relation to the emitting source.
Secondly, to efficiently excite a sensing element, such as the NV-containing diamond, the core and diamond diameters should be matched, or an optical structure needs to be attached.

\section{Conclusion}

In summary, we have reported a characterization of the background light generated in fibers relevant to NV-based fluorescence sensing applications. 
Two processes were found to dominate the measured spectra of the investigated fibers.
Below $\SI{600}{\nano\meter}$, it consists of Raman peaks that shift in their emission wavelength with the excitation wavelength and in their intensity with the excitation power.
Above $\SI{600}{\nano\meter}$, it is dominated by the NBOHC-related fluorescence band with a broad maximum at $\approx\SI{645}{\nano\meter}$.
Furthermore, an additional peak at $\approx\SI{695}{\nano\meter}$ is observed for small-core fibers and is also visible in the spectrum of an empty illuminated ceramic connector.
Additionally, the influence of the fiber assembly (epoxy, tubing, connector, polishing) and physical properties (fiber length, bending, coupling, cooling) was investigated.

From our findings, we extract practical guidelines for nitrogen-vacancy-center-based, fiber-coupled sensors. 
First, fiber length should be minimized, and low-background fibers, such as AFM200L, UM22-100, and FP200ERT, are preferred. 
As no monotonic dependence on core diameter or numerical aperture was observed, it is recommended to compare the spectra of available fibers before employing them in a sensor.
In general, ferrule illumination should be avoided. 
As an epoxy for the fiber assembly, Epo-Tek 353ND showed the lowest autofluorescence, but its influence is weak. 
Furthermore, the thermal properties should be controlled, as cooling can affect the autofluorescence intensity. 
Bending and minor coupling-angle variations do not show an impact. 
Also, thorough polishing and cleaning of the fiber facets can minimize the background intensity. 
Future work could quantify long-term drift (bleaching or aging) and include further analysis of fibers of the same type and different fiber core or cladding diameters.

\section{Acknowledgement}
This project was funded by the German Research Foundation (Deutsche Forschungsgemeinschaft DFG), Project-ID. 454931666, and the Quanten-Initiative Rheinland-Pfalz (QUIP).
Furthermore, we thank Robert Wendels for experimental support, and Gesa Welker for helpful discussions.

\section{Data availability}
All data shown in the images, plots and data supporting the findings in this manuscript is available on Zenodo~\cite{johansson_data_2026}.

\bibliography{literature}

\end{document}